\documentclass[12pt]{article}
\usepackage{amsfonts,amssymb}

\def\C{\mathbb{C}}
\def\N{\mathbb{N}}

\def\g{\mathfrak g}

\def\bq{ \begin{equation} }
\def\eq{ \end{equation} }
\def\ben{ \begin{eqnarray} }
\def\en{ \end{eqnarray} }

\def\frac#1#2{{#1\over #2}}

\def\on#1#2{\mathop{\vbox{\ialign{##\crcr\noalign{\kern2pt}
$\scriptstyle{#2}$\crcr\noalign{\kern2pt\nointerlineskip}
\kern-2pt$\hfil\displaystyle{#1}\hfil$\crcr}}}\limits}

\begin{document}

\baselineskip=15pt
\vspace{1cm} \centerline{{\LARGE \textbf {Integrable
pseudopotentials related to
 }}}
\vspace{0.3cm} \centerline{{\LARGE \textbf {generalized
hypergeometric functions
 }}}

\vskip1cm \hfill
\begin{minipage}{13.5cm}
\baselineskip=15pt {\bf A.V. Odesskii ${}^{1,}{}^{2}$,  V.V. Sokolov
${}^{1}$}
\\ [2ex] {\footnotesize
${}^{1}$  L.D. Landau Institute for Theoretical Physics (Russia)
\\
${}^{2}$  University of Manchester (UK)
\\}
\vskip1cm{\bf Abstract}

We construct integrable pseudopotentials with an arbitrary number of
fields in terms of generalized hypergeometric functions. These
pseudopotentials yield some integrable (2+1)-dimensional
hydrodynamic type systems. In two particular cases these systems are
equivalent to integrable scalar 3-dimensional equations of second
order. An interesting class of integrable (1+1)-dimensional
hydrodynamic type systems is also generated by our pseudopotentials.

\end{minipage}

\vglue1cm \textbf{Address}: L.D. Landau Institute for Theoretical
Physics of Russian Academy of Sciences, Kosygina 2, 119334,
Moscow, Russia

\textbf{E-mail}: alexander.odesskii@manchester.ac.uk,
odesskii@itp.ac.ru, sokolov@itp.ac.ru

\newpage \tableofcontents
\newpage

\section{Introduction}

The main object of integrability theory is the Lax equation
\begin{equation} \label{lax} L _{t}=[L,\,A]
.\end{equation} Here $A$ and $L$ are operators depending on
functions $u_1,..., u_n$ and (\ref{lax}) is equivalent to a system
of nonlinear differential equations for $u_i$. For the KP-hierarchy
and its different reductions $A$ is a linear differential operator
$A=\sum r_i \partial_x^i,$ whose coefficients $r_i$ are differential
polynomials in $u_1,..., u_n.$ The $L$-operator could be a
differential operator or a more complicated object like a ratio of
two differential operators or a formal (non-commutative) Laurent
series with respect to $\partial_x^{-1}.$

The dispersionless analog of (\ref{lax}) has the following form
\begin{equation} \label{maineq}L _{t}=\{L ,A
\},\end{equation} where $\{L ,A \}=A_{p}L _{x}-A_{x} L_{p}.$ As
usual, the commutator in (\ref{lax}) is replaced by the Poisson
bracket and the non-commutative variable $\partial_x$ by the
commutative "spectral"  parameter $p.$ The transformation
$L(x,t,p)\rightarrow p(x,t,L)$ reduces (\ref{maineq}) to the
following conservative form
\begin{equation}  \label{con}
p_t=A(p,u_{1},\dots, u_{n})_x,
\end{equation}
where $L$ plays the role of a parameter. The latter equation can be
rewritten as
\begin{equation}  \label{psi}
\psi_t=A(\psi_{x},u_{1},\dots, u_{n}),
\end{equation}
where $p=\psi_{x}$.

Equations (\ref{psi}) can be chosen as a basis, on which a theory of
integrable 3-dimensional dispersionless PDEs can be built.
 Most such equations can be written in the form
\begin{equation}   \label{genern}
\sum_{j=1}^n a_{ij}({\bf u})\,u_{j,t_{1}}+\sum_{j=1}^n
b_{ij}({\bf u})\,u_{j, t_{2}}+ \sum_{j=1}^n c_{ij}({\bf
u})\,u_{j, t_{3}}=0, \qquad i=1,...,l,
\end{equation}
where ${\bf u}=(u_{1},\dots, u_{n}).$ All known integrable systems
(\ref{genern}) admit the so-called pseudopotential representation
\begin{equation} \label{pseudo}
\psi_{t_{2}}=A(\psi_{t_{1}},{\bf u}), \qquad  \psi_{t_{3}}=B(\psi_{t_{1}},{\bf u}),
\end{equation}
by means of a pair of equations (\ref{psi}) whose the compatibility
conditions $\psi_{{t_2}t_{3}}=\psi_{t_{3}t_{2}}$ are equivalent to
(\ref{genern}). The functions $A,~B$ are called pseudopotentials.
Such a
 pseudopotential representation is a dispersionless version  of the zero curvature
representation, which is a basic notion in the integrability theory
of solitonic equations (see \cite{zahshab}) .

One of the interesting and attractive  features of the theory of
integrable dispersionless equations is that the dependence of the
pseudopotentials $A(p,u_{1},\dots,u_{n})$ on $p$ can be much more
complicated then in the solitonic case. For instance, in \cite{kr4,
dub} some important examples of pseudopotentials $A$ were found
related to the Whitham averaging procedure for integrable dispersion
PDEs and to the Frobenious manifolds. For these examples the
$p$-dependence is determined by an algebraic curve of arbitrary
genus $g$. In the paper \cite{odsok} a certain class of
pseudopotentials with movable singularities was described. Some of
the pseudopotentials constructed in \cite{odsok} are written in
terms of degenerate hypergeometric functions.

In the paper \cite{odes} a wide class of pseudopotentials
$A(p,u_1,...,u_n)$ related to rational algebraic curves was
constructed. These pseudopotentials were written in the following
parametric form:
$$A=F_1(\xi,u_1,...,u_n),\qquad p=F_2(\xi,u_1,...,u_n),$$
where the $\xi$-dependence of the functions $F_i$ is defined by the
ODE
\begin{equation}\label{ODE}
\label{par}F_{i,\xi}=\phi_i(\xi,u_1,...,u_n)\cdot\xi^{-s_1}(\xi-1)^{-s_2}(\xi-u_1)^{-s_3}...(\xi-u_n)^{-s_{n+2}}.
\end{equation}
Here $s_1,...,s_{n+2}$ are arbitrary constants and $\phi_i$ are
polynomials in $\xi$ of degree $n$. The dependence of
  $\phi_i$ on $u_{1},\dots,u_{n}$ was described in
terms of solutions of some overdetermined linear system of PDEs with
rational coefficients.

In this paper we generalize this result and construct new classes of
pseudopotentials \linebreak $A_{n,k}(p,u_1,...,u_n)$ whose $p$-dependence is
given by (\ref{ODE}), where $\phi_i(\xi)$ are polynomials in $\xi$ of
degree $n-k,\,$ $k=0,...,n-1$. We call the corresponding functions
$A_{n,k}$ {\it pseudopotentials of defect $k$}. The pseudopotentials
of defect 0 are just pseudopotentials from \cite{odes} written in a
different form.

We describe the pseudopotentials of defect $k$ in terms of linearly independent
solutions of the following system of linear PDEs with rational
coefficients
\begin{equation} \label{Darbu}
\frac{\partial^2 h}{\partial u_i \partial
u_j}=\frac{s_i}{u_i-u_j}\cdot \frac{\partial h}{\partial
u_j}+\frac{s_j}{u_j-u_i}\cdot \frac{\partial h}{\partial
u_i},~~~i,j=1,...,n, \qquad i\ne j,
\end{equation}
and
\begin{equation} \begin{array}{c} \label{Hyper}
\displaystyle \frac{\partial^2 h}{\partial u_i \partial
u_i}=-\Big(1+\sum_{j=1}^{n+2} s_j\Big) \frac{s_i}{u_i (u_i-1) }\, h+
\frac{s_i}{u_i (u_i-1)} \sum_{j\ne i}^n \frac{u_j
(u_j-1)}{u_j-u_i}\cdot
\frac{\partial h}{\partial u_j}+\\[7mm]
\displaystyle \Big(\sum_{j\ne i}^n \frac{s_j}{u_i-u_j}+
\frac{s_i+s_{n+1}}{u_i}+ \frac{s_i+s_{n+2}}{u_i-1}\Big)
\frac{\partial h}{\partial u_i}
\end{array}
\end{equation}
for one unknown function $h(u_1, \dots, u_n)$.  If $n=1$, then we
have no equations (\ref{Darbu}) and the single equation
(\ref{Hyper}) coincides with the standard hypergeometric equation,
$$u (u-1) \, h(u)''+ [(\alpha+\beta+1)\,u-\gamma]\, h(u)'+\alpha
\beta\, h(u)=0,$$ where $s_1=-\alpha, \,$ $s_2=\alpha-\gamma$,
$s_3=\gamma-\beta-1$. Notice, that hypergeometric functions already appeared in connection with
dispersionless PDEs (see, for example \cite{pav4,odsok, odpavsok}).
For arbitrary $n$ the system (\ref{Darbu}),
(\ref{Hyper}) can be solved in terms of {\it generalized
hypergeometric functions} (see \cite{trfun,gel}).

Note that the pseudopotential $A_{n,k}$ is written in terms of $k+2$
linearly independent solutions of the system (\ref{Darbu}),
(\ref{Hyper}) and therefore the group $GL_{k+2}$ acts on the set of
such pseudopotentials. If $k=0$, then this is just the usual action
of $GL_2$ on the space of independent variables $x,~t$ in the
equation (\ref{psi}). In the case $k>0$ the action of larger group
$GL_{k+2}$ is still to be explained. For a particular class of
3-dimensional equations the existence of such a group of symmetries
was pointed out in \cite{odfer} (see also \cite{burferts}). It is
known \cite{dorfer,odfer} that knowledge of the symmetry group
$GL_{n}$ allows us to linearize systems of ODEs and PDEs.

The paper is organized as follows.

In Section 2 we describe some properties of system (\ref{Darbu}),
(\ref{Hyper}) and its solutions needed for our purposes. Most of
these properties are well known to experts.

In Section 3 we rewrite formulas of the paper \cite{odes} in terms
of generalized hypergeometric functions. In our paper the
pseudopotentials constructed in \cite{odes} are called
pseudopotentials of defect 0. A couple of such pseudopotentials
defines a system of the form (\ref{genern}) with $l=n.$ These
systems are rewritten  in terms of generalized hypergeometric
functions in Section 3. We also prove that each of these systems
admits $n+1$ conservation laws of hydrodynamic type.

In Section 4, for any $n$ and $k>0$ we construct  {\it
pseudopotentials of defect $k$}. A couple of such pseudopotentials
defines a system of the form (\ref{genern}) with $l=n+k.$ These
systems are also constructed in Section 4. The particular cases $n=3,~ k=1$ and
$n=5,~ k=3$ yield integrable equations of the form
\begin{equation}   \label{aqq}
\sum_{i,j} P_{i,j} (z_{t_1},z_{t_2},z_{t_3})\, z_{t_i,t_j}=0, \qquad
i,j=1,2,3,
\end{equation}
and
\begin{equation}\label{hir}Q(z_{t_1,t_1},z_{t_1,t_2},z_{t_1,t_3},
z_{t_2,t_2},z_{t_2,t_3},z_{t_3,t_3})=0.\end{equation}
A classification of all integrable equations (\ref{aqq}) and (\ref{hir})
was presented in \cite{burferts} and in \cite{ferhadhus},
correspondingly. Our integrable equations give generic solutions of these classification problems.

In Sections 5 we construct and study a certain class of integrable
(1+1)-dimensional hydrodynamic type systems of the form
\begin{equation}\label{gidra}
r_{t}^{i}=v^{i}(r^1,...,r^N)r_{x}^{i},\qquad i=1,2,...,N.
\end{equation}
These systems are defined by an
universal overdetermined compatible system of PDEs of the
Gibbons-Tsarev type \cite{Gibt, pav2} for some functions
$w(r^1,...,r^N)$,$~\xi_1(r^1,...,r^N),...,$ $\xi_N(r^1,...,r^N).$
This system has the following form
\begin{equation}\label{gibtsar1}
\partial_i\xi_j=\frac{\xi_j(\xi_j-1)}{\xi_i-\xi_j}\partial_i w,~~~
\partial_{ij}w=\frac{2\xi_i\xi_j-\xi_i-\xi_j}{(\xi_i-\xi_j)^2}\partial_iw\partial_jw,\qquad i,j=1,...,N,~~~i\ne
j.
\end{equation}
The only velocities $v^{i}(r^1,...,r^N)$ in (\ref{gidra}) depend on
$n$, $k.$ They are described by $k+2$ linearly independent solutions
of the linear system (\ref{Darbu}), (\ref{Hyper}) (see Section 5,
Theorem 3). One has to substitute functions
$u_1=u_1(r^1,...,r^N),...,u_n=u_n(r^1,...,r^N)$ for the arguments of
these solutions. The functions $u_{i}$ are also universal. They are
defined by the following system of PDEs
\begin{equation}\label{u1}
\partial_iu_j=\frac{u_j(u_j-1)\partial_iw}{\xi_i-u_j},~~~i=1,...,N,~~~j=1,...,n.
\end{equation}
It is easy to verify that the system (\ref{gibtsar1}), (\ref{u1}) is
consistent. Therefore our (1+1)-dimensional systems (\ref{gidra})
admit a local parametrization by $2 N$ functions of one variable.

For some very special values of parameters $s_{i}$ in (\ref{Darbu}),
(\ref{Hyper}) our systems (\ref{gidra}) are related to the Whitham
hierarchies \cite{kr4}, to the Frobenious manifolds \cite{dub,kr},
and to the associativity equation \cite{dub,kr}.

In Section 6 we recall the definition of hydrodynamic reductions.
According to \cite{ferhus1}, the existence of sufficiently many
hydrodynamic reductions can be chosen as a definition of the
integrability of the systems (\ref{genern}). We also recall the
definition of integrable pseudopotentials (see \cite{odpavsok}). We
introduce the notion of compatible pseudopotentials and notice that
each pair of them gives
 a system (\ref{genern}) that admits both a
pseudopotential representation and sufficiently many hydrodynamic
reductions. We show that the (1+1)-dimensional hydrodynamic type
systems found in Section 5 are hydrodynamic reductions of our
pseudopotentials $A_{n,k}$. This implies that these pseudopotentials and
the corresponding 3-dimensional systems
are integrable in the sense of the definitions mentioned above (see Theorem 4).

\section{Generalized hypergeometric functions}

The following statements can be verified straightforwardly.

{\bf Proposition 1.} The system of linear equations (\ref{Darbu}),
(\ref{Hyper}) is compatible for any constants $s_1,\dots, s_{n+2}.$ The
dimension of  the linear space  ${\cal H}$ of
solutions of the system (\ref{Darbu}), (\ref{Hyper}) is equal to
$n+1$. $\blacksquare$

We call elements of  ${\cal H}$ {\it generalized hypergeometric functions.}

{\bf Proposition 2.} The system (\ref{Darbu}), (\ref{Hyper}) is
equivalent to the following system
\begin{equation}\label{hypgen}
Q_i(u_1\frac{\partial}{\partial u_1},...,u_n\frac{\partial}{\partial
u_n})u_i^{-1}h=P_i(u_1\frac{\partial}{\partial
u_1},...,u_n\frac{\partial}{\partial u_n})h,~~~i=1,...,n
\end{equation}
where
$$Q_i(k_1,...,k_n)=(k_1+...+k_n-s_1-...-s_{n+1})(k_i+1),$$
$$P_i(k_1,...,k_n)=(k_1+...+k_n-1-s_1-...-s_{n+2})(k_i-s_i).$$
$\blacksquare$

Recall that a system of the form (\ref{hypgen}) is called a
hypergeometric system \cite{gel}. It can be solved in terms of the
so-called Horn series \cite{gel}.

{\bf Example 1.} The system (\ref{hypgen}) and hence (\ref{Darbu}),
(\ref{Hyper}) has a unique solution  holomorphic at the point ${\bf
0}=(0,\dots, 0)$ such that $h({\bf 0})=1.$ The derivatives of this
solution at ${\bf 0}$ is given by
$$
h^{(k_1,\dots,k_n)}({\bf 0})=\frac{\prod_{j=0}^{k_1+\dots+k_n-1}
 (1-j+ s_{n+2}+r)}{\prod_{j=0}^{k_1+\dots+k_n-1} (-j+r)} \,
\prod_{j=1}^{n} \prod_{i=0}^{k_j-1} (i-s_j),
$$
where
$$
r=\sum_{i=1}^{n+1} s_i.$$

Let us denote the solution described in this example by
$F(s_1,\dots, s_{n+2}, \, u_1,\dots, u_n)$. For brevity, we also
will use the notation $F(s_1,\dots, s_{n+2} )$.

{\bf Proposition 3.} The function $F(s_1,\dots, s_{n+2} )$ admits
the following integral representation
$$
F(s_1,\dots, s_{n+2}, \, u_1,\dots, u_n)= C \int_0^1
t^{-2-r-s_{n+2}} (1-t)^{s_{n+2}} (1-t u_1)^{s_1}\cdots (1-t
u_n)^{s_n} \,dt,
$$
where
$$
C=\frac{\Gamma (-r)}{\Gamma (1+s_{n+2}) \Gamma (-1-r-s_{n+2})}.
$$ $\blacksquare$

It is well-known that for the standard hypergeometric equation there
exist the Laplace transformations shifting the parameters by 1.
Analogies of such transformations for the system (\ref{Darbu}),
(\ref{Hyper}) are given by

{\bf Proposition 4.} The following identities hold:
$$
\frac{\partial F(s_1,\dots,s_i,\dots s_{n+2})
 }{\partial u_i}= -\frac{s_i (1+r+s_{n+2})}{r}\,
F(s_1,\dots,s_i-1,\dots s_{n+2}),\qquad i\le n,
$$
$$
L_1 \Big(F(s_1,\dots,s_{n+1}, s_{n+2})\Big)  = \frac{s_{n+1}
(1+r+s_{n+2})}{r}\, F(s_1,\dots,s_{n+1}-1, s_{n+2}),
$$
$$
L_2 \Big(F(s_1,\dots,s_{n+1}, s_{n+2})\Big)  =  (1+r+s_{n+2}) \,
F(s_1,\dots,s_{n+1}, s_{n+2}-1),
$$
where
$$
L_1=\sum_{j=1}^n (1-u_j) \frac{\partial}{\partial
u_j}+(1+r+s_{n+2}), \qquad L_2=-\sum_{j=1}^n u_j
\frac{\partial}{\partial u_j}+(1+r+s_{n+2}),
$$
and
$$
M_i \Big(F(s_1,\dots,s_i,\dots s_{n+2})\Big)  =  (1+r) \,
F(s_1,\dots,s_i+1,\dots s_{n+2}),\qquad i\le n,
$$
$$
M_{n+1} \Big(F(s_1,\dots,s_{n+1}, s_{n+2})\Big)  = (1+r)\,
F(s_1,\dots,s_{n+1}+1, s_{n+2}),
$$
$$
M_{n+2} \Big(F(s_1,\dots,s_{n+1}, s_{n+2})\Big)  =  -(1+s_{n+2}) \,
F(s_1,\dots,s_{n+1}, s_{n+2}+1),
$$
where
$$
M_i=\sum_{j=1}^n u_j (u_j-1) \frac{\partial}{\partial
u_j}-\sum_{j=1}^n s_j u_j-(2+r+s_{n+2}) u_i
 +(1+r), \qquad i\le n,
$$
$$
M_{n+1}=\sum_{j=1}^n u_j (u_j-1) \frac{\partial}{\partial
u_j}-\sum_{j=1}^n s_j u_j
 +(1+r),
$$
$$
M_{n+2}=\sum_{j=1}^n u_j (u_j-1) \frac{\partial}{\partial
u_j}-\sum_{j=1}^n s_j u_j
 -(1+s_{n+2}).
$$
Furthermore, let ${\cal H}_{s_1,...,s_{n+2}}$ be the space of solutions
of the system (\ref{Darbu}), (\ref{Hyper}). We have
$$\frac{\partial}{\partial u_i}{\cal H}_{s_1,...,s_{n+2}}\subset
{\cal H}_{s_1,...,s_i-1,...,s_{n+2}},~~~L_1{\cal H}_{s_1,...,s_{n+2}}\subset
{\cal H}_{s_1,...,s_{n+1}-1,s_{n+2}},$$
$$
 L_2{\cal H}_{s_1,...,s_{n+2}}\subset
{\cal H}_{s_1,...,S_{n+2}-1}, \qquad
M_i{\cal H}_{s_1,...,s_{n+2}}\subset
{\cal H}_{s_1,...,s_i+1,...,s_{n+2}}.$$  $\blacksquare$

{\bf Proposition 5.} Let ${\cal H}={\cal H}_{s_1,...,s_{n+2}}$ and
$\widetilde{{\cal H}}={\cal H}_{s_1,...,s_n,0,s_{n+1},s_{n+2}}$.
Then $\widetilde{{\cal H}}$ is spanned by ${\cal H}$ and the
function
\begin{equation}\label{gen}
Z(u_1,...,u_n,u_{n+1})=\int_0^{u_{n+1}}(t-u_1)^{s_1}\cdots(t-u_n)^{s_n}t^{s_{n+1}}(t-1)^{s_{n+2}}dt.
\end{equation}
Moreover, the space ${\cal H}_{s_1,...,s_n,0,...,0,s_{n+1},s_{n+2}}$
($m$ zeros) is spanned by ${\cal H}$ and $Z(u_1,...,u_n,u_{n+1})$,
$Z(u_1,...,u_n,u_{n+2}), ..., Z(u_1,...,u_n,u_{n+m})$.
$\blacksquare$

\section{Pseudopotentials of defect 0}

Most results of this Section was obtained in a different form in the
paper \cite{odes}.

For any generalized hypergeometric function $g\in\cal H$ we put
\begin{equation}\label{pol0}
S_n(g, \xi)=\sum_{1\leq i\leq
n}u_i(u_i-1)(\xi-u_1)...\hat{i}...(\xi-u_n)\, g_{u_i}+(1+\sum_{1\leq
i\leq n+2}s_i)(\xi-u_1)...(\xi-u_n)\,g.
\end{equation}
Here $\displaystyle g_{u_i}=\frac{\partial g}{\partial u_{i}}.$ It
is clear that $S_n(g, \xi)$ is a polynomial of degree $n$ in $\xi.$

{\bf Example 2.} In the simplest case $n=1$ we have
$$S_1(g,\xi)=u(u-1)g_u+(1+s_1+s_2+s_3)(\xi-u)g$$
where $u=u_1$.

We need the following property of the polynomial $S_n(g,\xi)$:

{\bf Lemma 1.} For any $1\le m \le n$
the following identity is valid
$$\displaystyle u_m(u_m-1)(u_m-u_1)...\hat{m}...(u_m-u_n)\frac{S_n(g,\xi)_{u_m}+\frac{(s_m+1)S_n(g,\xi)}{\xi-u_m}}{S_n(g,u_m)}=$$
$$\xi(\xi-1)(\xi-u_1)...\hat{m}...(\xi-u_n)\Big(\frac{s_1}{\xi-u_1}+...+\frac{s_m+1}{\xi-u_m}+...+\frac{s_n}{\xi-u_n}
+\frac{s_{n+1}}{\xi}+ \frac{s_{n+2}}{\xi-1}\Big).$$  $\blacksquare$

Define $P_n(g,\xi)$ by the formula
\begin{equation}\label{pp}P_n(g,\xi)=\int_0^{\xi}S_n(g,\xi)(\xi-u_1)^{-s_1-1}...(\xi-u_n)^{-s_n-1}
\xi^{-s_{n+1}-1}(\xi-1)^{-s_{n+2}-1}d\xi\end{equation} if ${\rm Re}\, s_{n+1}<-1$ and
as the analytic continuation of this expression otherwise.

{\bf Proposition 6.} The expression
\begin{equation}\label{pr1}
\frac{P_n(g,\xi)_{u_m}}{S_n(g,u_m)}
\end{equation}
does not depend on $g$. More precisely,
\begin{equation}\begin{array}{c}
\label{pr2}
\displaystyle u_m(u_m-1)(u_m-u_1)...\hat{m}...(u_m-u_n)\frac{P_n(g,\xi)_{u_m}}{S_n(g,u_m)}=\\[5mm]
-(\xi-u_1)^{-s_1}...(\xi-u_m)^{-s_m-1}...(\xi-u_n)^{-s_n}
\xi^{-s_{n+1}}(\xi-1)^{-s_{n+2}}.
\end{array}
\end{equation}

{\bf Proof.} The derivative of (\ref{pr1}) with respect to $\xi$ is
equal to
$$\displaystyle \frac{S_n(g,\xi)_{u_m}+\frac{(s_m+1)S_n(g,\xi)}{\xi-u_m}}{S_n(g,u_m)}
(\xi-u_1)^{-s_1-1}...(\xi-u_n)^{-s_n-1}
\xi^{-s_{n+1}-1}(\xi-1)^{-s_{n+2}-1}.$$ Lemma 1 implies that this
derivative does not depend on $g$. Since the value of (\ref{pr1}) at
$\xi=0$ is equal to zero, expression (\ref{pr1}) itself does not
depend of $g$. Identity (\ref{pr2}) also follows from Lemma 1.
$\blacksquare$

Let $g_1,~g_0$ be linearly independent elements of ${\cal H}$. A
pseudopotential $A_n(p,u_1,...,u_n)$ defined in a parametric form by
\begin{equation}\label{psdef0}
A_n=P_n(g_1,\xi), \qquad p=P_n(g_0,\xi)
\end{equation}
is called {\it pseudopotential of defect 0}.

Relations (\ref{psdef0}) mean that to find $A_{n}(p,u_1,...,u_n),$ we have to express $\xi$ from the second equation and
substitute the result into the first equation.

Let $g_0,g_1,...,g_n\in {\cal H}$ be a basis in ${\cal H}.$ Define
pseudopotentials $B_{\alpha}(p,u_1,...,u_n)$  of defect 0, where
$\alpha=1,...,n,$ by
$$B_{\alpha}=P_n(g_{\alpha},\xi), \qquad p=P_n(g_0,\xi), \qquad \alpha=1,...,n.$$

Suppose that $u_1,...,u_n$ are functions
of $t_0=x,~t_1,...,t_n$.

{\bf Theorem 1.}
The compatibility conditions
$\psi_{t_{\alpha}t_{\beta}}=\psi_{t_{\beta}t_{\alpha}}$ for the
system
\begin{equation}
\label{pseudopar}\psi_{t_{\alpha}}=B_{\alpha}(\psi_x,u_1,...,u_n),
\qquad \alpha=1,...,n.
\end{equation}
are equivalent to the following system of PDEs for $u_1,...,u_n$:
$$\sum_{1\leq i\leq n,i\ne
j}(g_{q,u_j}g_{r,u_i}-g_{r,u_j}g_{q,u_i})\frac{u_j(u_j-1)u_{i,t_{s}}-u_i(u_i-1)u_{j,t_{s}}}{u_j-u_i}+
\sigma\cdot (g_{q}g_{r,u_j}-g_{r}g_{q,u_j})u_{j,t_{s}}+$$
\begin{equation}
\label{3dsyst} \sum_{1\leq i\leq n,i\ne
j}(g_{r,u_j}g_{s,u_i}-g_{s,u_j}g_{r,u_i})\frac{u_j(u_j-1)u_{i,t_{q}}-u_i(u_i-1)u_{j,t_{q}}}{u_j-u_i}+
\sigma\cdot (g_{r}g_{s,u_j}-g_{s}g_{r,u_j})u_{j,t_{q}}+
\end{equation}
$$\sum_{1\leq i\leq n,i\ne
j}(g_{s,u_j}g_{q,u_i}-g_{q,u_j}g_{s,u_i})\frac{u_j(u_j-1)u_{i,t_{r}}-u_i(u_i-1)u_{j,t_{r}}}{u_j-u_i}+
\sigma\cdot(g_{s}g_{q,u_j}-g_{q}g_{s,u_j})u_{j,t_{r}}=0,$$ where
$j=1,...,n,\,$ $\sigma=1+s_1+...+s_{n+2}$. Here $q, r, s$ run from 0
to $n$ and $t_0=x$.

{\bf Proof.} If $B_{\alpha}$ are given in a parametric form
$$B_{\alpha}=f_{\alpha}(\xi,u_1,...,u_n),\qquad p=f_0(\xi,u_1,...,u_n),$$
then the compatibility conditions for (\ref{pseudopar}) is equivalent to
\begin{equation}
\label{whit}
\sum_{i=1}^n\Big((f_{q,\xi}f_{r,u_i}-f_{r,\xi}f_{q,u_i})u_{i,t_{s}}+
(f_{r,\xi}f_{s,u_i}-f_{s,\xi}f_{r,u_i})u_{i,t_{q}}+
(f_{s,\xi}f_{q,u_i}-f_{q,\xi}f_{s,u_i})u_{i,t_{r}}\Big)=0.
\end{equation}

Taking into account (\ref{pp}), (\ref{pr2}), we get
$$
\begin{array}{c}
f_{q,\xi}f_{r,u_i}-f_{r,\xi}f_{q,u_i}=\\[5mm]
\Big(S_n(g_{q},\xi)P_n(g_{r},\xi)_{u_i}-
S_n(g_{r},\xi)P_n(g_{q},\xi)_{u_i}\Big)(\xi-u_1)^{-s_1-1}\cdots
(\xi-u_n)^{-s_n-1}\xi^{-s_{n+1}-1}(\xi-1)^{-s_{n+2}-1}\\[5mm]
\displaystyle =\frac{S_n(g_{q},\xi)S_n(g_{r},u_i)-
S_n(g_{r},\xi)S_n(g_{q},u_i)}{(\xi-u_i)\cdot
u_i(u_i-1)(u_i-u_1)...\hat{i}...(u_i-u_n)}\cdot
T=\frac{S_n(g_{q},\xi)g_{r,u_i}-
S_n(g_{r},\xi)g_{q,u_i}}{\xi-u_i}\cdot T.
\end{array}$$
 Here
$$
T=-(\xi-u_1)^{-2s_1-1}...
(\xi-u_n)^{-2s_n-1}\xi^{-2s_{n+1}-1}(\xi-1)^{-2s_{n+2}-1}
$$
does not depend on $i$. Using the above formula for $f_{q,\xi}f_{r,u_i}-f_{r,\xi}f_{q,u_i}$ and
similar formulas for
$f_{r,\xi}f_{s,u_i}-f_{s,\xi}f_{r,u_i},~
f_{s,\xi}f_{q,u_i}-f_{q,\xi}f_{s,u_i}$,
we can rewrite (\ref{whit}) as follows:
$$\sum_{1\leq i\leq n}\Big(\frac{S_n(g_{q},\xi)g_{r,u_i}-S_n(g_{r},\xi)g_{q,u_i}}{\xi-u_i}u_{i,t_{s}}+
\frac{S_n(g_{r},\xi)g_{s,u_i}-S_n(g_{s},\xi)g_{r,u_i}}{\xi-u_i}u_{i,t_{q}}+$$$$
\frac{S_n(g_{s},\xi)g_{q,u_i}-S_n(g_{q},\xi)g_{s,u_i}}{\xi-u_i}u_{i,t_{r}}\Big)=0.$$
It follows from (\ref{pol0}) that the left hand side is a polynomial
of degree $n-1$ in $\xi$. To conclude the proof, it remains to
evaluate this polynomial at $\xi=u_1,...,u_n$.  $\blacksquare$

{\bf Remark 1.} Given $t_{1},t_{2},t_{3},$ Theorem 1 yields a
3-dimensional system of the form (\ref{genern}) with $l=n$ equations
possessing pseudopotential representation.

{\bf Remark 2.} A system of PDEs for $u_1,...,u_n,$ which is
equivalent to compatibility conditions for equations of the form
(\ref{whit}), was called  in \cite{kr4} a {\it Whitham hierarchy}.
In the paper \cite{kr4} I.M. Krichever constructed some Whitham
hierarchies related to algebraic curves of arbitrary genus $g$. The
hierarchy corresponding to $g=0$ is equivalent to one described by
Theorem 1 if $s_{1}=\dots=s_{n+2}=0$. In this case the vector space
${\cal H}$ is spanned by $1,u_{1}, u_{2},\dots, u_{n}$.

{\bf Proposition 7.} The system (\ref{3dsyst}) possesses $n+1$ hydrodynamic type conservation laws.

{\bf Proof.} Let $\widehat{\cal H}={\cal
H}_{-2s_1,...,-2s_n,-2s_{n+1}-1,-2s_{n+2}-1}$ be the space of
generalized hypergeometric functions defined by (\ref{Darbu}),
(\ref{Hyper}) with $\hat{s}_i=-2s_i$ for $i=1,...,n$ and
$\hat{s}_i=-2s_i-1$ for $i=n+1,~n+2.$ Let $Z\in\widehat{\cal H}$ be
an arbitrary element in $\widehat{\cal H}$. Denote by $X_j$ the left
hand side of (\ref{3dsyst}). Define functions $A_i,B_i,C_i$ by
$$\sum_{i=1}^n(A_iu_{i,t_q}+B_iu_{i,t_r}+C_iu_{i,t_s})=\sum_{j=1}^n\frac{1}{s_j}Z_{u_j}X_j.$$
One can check that $(A_i)_{u_j}=(A_j)_{u_i}$,
$(B_i)_{u_j}=(B_j)_{u_i}$, $(C_i)_{u_j}=(C_j)_{u_i}$. Therefore
$A_i=A_{u_i}$, $B_i=B_{u_i}$, $C_i=C_{u_i}$ for some functions
$A,~B,~C$ and we have
$$A_{t_q}+B_{t_r}+C_{t_s}=0.$$
Since $\dim\widehat{\cal H}=n+1,$ we obtain $n+1$
conservation laws of the hydrodynamic type.

\section{Pseudopotentials of defect $k>0$}

In this section we construct a new class of pseudopotentials. We
call them {\it pseudopotentials of defect} $k.$ To define
pseudopotentials of defect $k,$ we fix $k$ linearly independent
generalized hypergeometric functions $h_1,...,h_k\in {\cal H}$. For
any $g\in {\cal H}$ define $S_{n,k}(g,\xi)$ by the formula
\begin{equation}\label{polgen}
S_{n,k}(g,\xi)=\frac{1}{\Delta}\sum_{1\leq i\leq
n-k+1}u_i(u_i-1)(\xi-u_1)...\hat{i}...(\xi-u_{n-k+1})\Delta_i(g).
\end{equation}
Here
$$\Delta=\det\left(\begin{array}{ccc}h_1&...&h_k\\h_{1,u_{n-k+2}}&...&h_{k,u_{n-k+2}}
\\.........&...&.........\\h_{1,u_n}&...&h_{k,u_n}
\end{array}\right),\qquad \Delta_i(g)=\det\left(\begin{array}{cccc}g&h_1&...&h_k\\g_{u_i}&h_{1,u_i}&...&h_{k,u_i}
\\g_{u_{n-k+2}}&h_{1,u_{n-k+2}}&...&h_{k,u_{n-k+2}}
\\.........&...&...&.........\\g_{u_n}&h_{1,u_n}&...&h_{k,u_n}
\end{array}\right)\,.$$
It is  clear that  $S_{n,k}(g,\xi)$ is a polynomial in $\xi$ of
degree $n-k.$ Notice that $S_{n,k}(h_1,\xi)=...=S_{n,k}(h_k,\xi)=0$.
It is  easy to see that linear transformations $h_i\to
c_{i1}h_1+...+c_{ik}h_k$, $g\to g+d_1h_1+...+d_kh_k$ with constant
coefficients $c_{ij}$, $d_i$ do not change $S_{n,k}(g,\xi)$.

{\bf Example 3.} In the simplest case $n=2,~k=1$ we have
$$S_{2,1}(g,\xi)=u_1(u_1-1)(\xi-u_2)\frac{gh_{1,u_1}-g_{u_1}h_1}{h_1}+u_2(u_2-1)(\xi-u_1)\frac{gh_{1,u_2}-g_{u_2}h_1}{h_1}.$$

{\bf Lemma 2.} If $1\leq m<n-k+2,$ then the following identity is valid:
$$u_m(u_m-1)(u_m-u_1)...\hat{m}...(u_m-u_n)\frac{S_{n,k}(g,\xi)_{u_m}+\frac{(s_m+1)S_{n,k}(g,\xi)}{\xi-u_m}}{S_{n,k}(g,u_m)}=$$
$$-(u_m-u_{n-k+2})...(u_m-u_n)\frac{1}{\Delta}\sum_{1\leq i\leq
n-k+1}u_i(u_i-1)(\xi-u_1)...\hat{i}...(\xi-u_{n-k+1})\widetilde{\Delta_i}+$$
$$\frac{1}{\Delta}\sum_{n-k+2\leq i\leq n,1\leq j\leq n-k+1}(u_m-u_{n-k+2})...\hat{i}...(u_m-u_n)s_iu_j(u_j-1)
(\xi-u_1)...\hat{j}...(\xi-u_{n-k+1})\widetilde{\Delta_{i,j}}+$$
$$\frac{(s_m+1)u_m(u_m-1)(u_m-u_{n-k+2})...(u_m-u_n)(\xi-u_1)...\hat{m}...(\xi-u_{n-k+1})}{\xi-u_m}+$$
$$(u_m-u_{n-k+2})...(u_m-u_n)\sum_{1\leq i\leq n-k+1,i\ne
m}s_iu_i(u_i-1)\prod_{1\leq j\leq n-k+1,j\ne i,m}(\xi-u_j)+$$
$$(u_m-u_{n-k+2})...(u_m-u_n)(\xi-u_1)...\hat{m}...(\xi-u_{n-k+1})(\sum_{1\leq i\leq n-k+1}(u_m+u_i-1)s_i+2u_m-1)+$$
$$u_m(u_m-1)(\xi-u_1)...\hat{m}...(\xi-u_{n-k+1})\sum_{n-k+2\leq i\leq n}(u_m-u_{n-k+2})...\hat{i}...(u_m-u_n)s_i+$$
$$(u_m-u_{n-k+2})...(u_m-u_n)(\xi-u_1)...\hat{m}...(\xi-u_{n-k+1})((u_m-1)s_{n+1}+u_ms_{n+2}).$$
If $n-k+2\leq m,$ then
$$u_m(u_m-1)(u_m-u_1)...\hat{m}...(u_m-u_n)\frac{S_{n,k}(g)(\xi)_{u_m}+\frac{s_mS_{n,k}(g,\xi)}{\xi-u_m}}{S_{n,k}(g,u_m)}=$$
$$\frac{1}{\Delta}\sum_{n-k+2\leq i\leq n,1\leq j\leq n-k+1}(u_m-u_{n-k+2})...\hat{i}...(u_m-u_n)s_iu_j(u_j-1)
(\xi-u_1)...\hat{j}...(\xi-u_{n-k+1})\widetilde{\Delta_{i,j}}+$$
$$\frac{s_mu_m(u_m-1)(u_m-u_{n-k+2})...\hat{m}...(u_m-u_n)(\xi-u_1)...(\xi-u_{n-k+1})}{\xi-u_m}.$$
Here
$$\widetilde{\Delta_i}=\det\left(\begin{array}{ccc}h_{1,u_i}&...&h_{k,u_i}\\h_{1,u_{n-k+2}}&...&h_{k,u_{n-k+2}}
\\.........&...&.........\\h_{1,u_n}&...&h_{k,u_n}
\end{array}\right)$$
and $\widetilde{\Delta_{i,j}}$ is obtained from $\Delta$ by
replacing the row $(h_{1,u_i},...,h_{k,u_i})$ by
$(h_{1,u_j},...,h_{k,u_j})$. $\blacksquare$

Define functions $P_{n,k}(g,\xi)$ by
\begin{equation}\label{pp2} P_{n,k}(g,\xi)=\end{equation}
$$\int_0^{\xi}S_{n,k}(g,\xi)(\xi-u_1)^{-s_1-1}...(\xi-u_{n-k+1})^{-s_{n-k+1}-1}
(\xi-u_{n-k+2})^{-s_{n-k+2}}...(\xi-u_n)^{-s_n}\xi^{-s_{n+1}-1}(\xi-1)^{-s_{n+2}-1}d\xi$$
if ${\rm Re}\, s_{n+1}<-1,$ and as the analytic continuation of this expression
otherwise.

{\bf Proposition 8.} The expression
\begin{equation}\label{pr4}
\frac{P_{n,k}(g,\xi)_{u_m}}{S_{n,k}(g,u_m)}
\end{equation}
does not depend on $g$. Moreover, we have
\begin{equation}\label{pr3}
\sum_{1\leq i\leq
k+1}u_{m_i}(u_{m_i}-1)(u_{m_i}-u_1)...\widehat{m_1,...,m_{k+1}}...(u_{m_i}-u_n)
\frac{P_{n,k}(g,\xi)_{u_{m_i}}}{S_{n,k}(g,u_{m_i})}=
\end{equation}
$$-\frac{(\xi-u_1)^{-s_1}...(\xi-u_{n-k+1})^{-s_{n-k+1}}
(\xi-u_{n-k+2})^{-s_{n-k+2}+1}...(\xi-u_n)^{-s_n+1}\xi^{-s_{n+1}}(\xi-1)^{-s_{n+2}}}{(\xi-u_{m_1})...(\xi-u_{m_{k+1}})}.$$

{\bf Proof.} The derivative of expression (\ref{pr4}) with respect to $\xi$ is
equal to
$$\frac{S_{n,k}(g,\xi)_{u_m}+\frac{(s_m+1)S_{n,k}(g,\xi)}{\xi-u_m}}{S_{n,k}(g,u_m)}(\xi-u_1)^{-s_1-1}\cdots$$$$(\xi-u_{n-k+1})^{-s_{n-k+1}-1}
(\xi-u_{n-k+2})^{-s_{n-k+2}}...(\xi-u_n)^{-s_n}\xi^{-s_{n+1}-1}(\xi-1)^{-s_{n+2}-1}$$
for $1\leq m<n-k+2$ and is equal to
$$\frac{S_{n,k}(g,\xi)_{u_m}+\frac{s_mS_{n,k}(g,\xi)}{\xi-u_m}}{S_{n,k}(g,u_m)}(\xi-u_1)^{-s_1-1}\cdots$$$$(\xi-u_{n-k+1})^{-s_{n-k+1}-1}
(\xi-u_{n-k+2})^{-s_{n-k+2}}...(\xi-u_n)^{-s_n}\xi^{-s_{n+1}-1}(\xi-1)^{-s_{n+2}-1}$$
otherwise. Lemma 2 implies that this derivative does not depend on
$g$. Moreover, the value of the expression (\ref{pr4}) at $\xi=0$ is
equal to zero. Therefore the expression (\ref{pr4}) itself does not
depend on $g$. The proof of (\ref{pr3}) is similar.  $\blacksquare$

Let $g_1,~g_2\in {\cal H}$. Assume that $g_1,g_2,h_1,...,h_k$ are
linearly independent. Define pseudopotential
$A_{n,k}(p,u_1,...,u_n)$ in parametric form by
\begin{equation}\label{psdef}
A_{n,k}=P_{n,k}(g_1,\xi),\qquad p=P_{n,k}(g_2,\xi).
\end{equation}
To construct  $A_{n,k}(p,u_1,...,u_n)$, we find $\xi$ from the second equation and
substitute into the first one.
The pseudopotential $A_{n,k}(p,u_1,...,u_n)$ is called {\it  pseudopotential of defect} $k.$

{\bf Theorem 2.} Let $g_0,g_1,...,g_{n-k},h_1,...,h_k\in {\cal H}$
be a basis in ${\cal H}$ and $B_{\alpha},~\alpha=1,...,n-k$ are
defined by
$$B_{\alpha}=P_{n,k}(g_{\alpha},\xi),\qquad p=P_{n,k}(g_0,\xi), \qquad \alpha=1,...,n-k.$$
Then the compatibility conditions for (\ref{pseudopar}) are
equivalent to the following system of PDEs for $u_1,...,u_n$:
$$
\sum_{1\leq i\leq n-k,i\ne
j}\Big(\Delta_j(g_q)\Delta_i(g_r)-\Delta_j(g_r)\Delta_i(g_q)\Big)\frac{u_j(u_j-1)u_{i,t_s}-u_i(u_i-1)u_{j,t_s}}{u_j-u_i}+
$$
\begin{equation}\label{syst}
\sum_{1\leq i\leq n-k,i\ne
j}\Big(\Delta_j(g_r)\Delta_i(g_s)-\Delta_j(g_s)\Delta_i(g_r)\Big)\frac{u_j(u_j-1)u_{i,t_q}-u_i(u_i-1)u_{j,t_q}}{u_j-u_i}+
\end{equation}
$$
\sum_{1\leq i\leq n-k,i\ne
j}\Big(\Delta_j(g_s)\Delta_i(g_q)-\Delta_j(g_q)\Delta_i(g_s)\Big)\frac{u_j(u_j-1)u_{i,t_r}-u_i(u_i-1)u_{j,t_r}}{u_j-u_i}=0,
$$
where $j=1,...,n-k$ and
\begin{equation}\label{syst1}
\sum_{i=1}^{n-k+1}\Delta_i(g_r)u_{i,t_s}=\sum_{i=1}^{n-k+1}\Delta_i(g_s)u_{i,t_r},
\end{equation}
\begin{equation}\label{syst2}
\sum_{i=1}^{n-k+1}\Delta_i(g_r)\frac{u_m(u_m-1)u_{i,t_s}-u_i(u_i-1)u_{m,t_s}}{u_m-u_i}=
\sum_{i=1}^{n-k+1}\Delta_i(g_s)\frac{u_m(u_m-1)u_{i,t_r}-u_i(u_i-1)u_{m,t_r}}{u_m-u_i},
\end{equation}
where $m=n-k+2,...,n$. Here $q, r, s$ run from 0 to $n$ and $t_0=x$.

{\bf Proof.} We have to explicitly calculate the coefficients in
(\ref{whit}).   Using (\ref{pp2}), (\ref{pr4}), we find that
$$f_{q,\xi}f_{r,u_i}-f_{r,\xi}f_{q,u_i}=\Big(S_{n,k}(g_{q},\xi)P_{n,k}(g_{r},\xi)_{u_i}-
S_{n,k}(g_{r},\xi)P_{n,k}(g_{q},\xi)_{u_i}\Big)\cdot
T=$$$$\Big(S_{n,k}(g_{q},\xi)S_{n,k}(g_{r},u_i)-
S_{n,k}(g_{r},\xi)S_{n,k}(g_{q},u_i)\Big)\cdot\frac{P_{n,k}(g_{q},\xi)_{u_i}}{S_{n,k}(g_{q},u_i))}\cdot
T.$$ Similar formulas are valid for
$f_{r,\xi}f_{s,u_i}-f_{s,\xi}f_{r,u_i},~
f_{s,\xi}f_{q,u_i}-f_{q,\xi}f_{s,u_i}$. Here
$$
T=(\xi-u_1)^{-s_1-1}...(\xi-u_{n-k+1})^{-s_{n-k+1}-1}
(\xi-u_{n-k+2})^{-s_{n-k+2}}...(\xi-u_n)^{-s_n}\xi^{-s_{n+1}-1}(\xi-1)^{-s_{n+2}-1}
$$
does not depend on $i$. Using (\ref{pr3}), we can express
$\displaystyle \frac{P_{n,k}(g_{q},\xi)_{u_i}}{S_{n,k}(g_{q},u_i)}$,
$i=1,...,n-k$ in terms of $\displaystyle
\frac{P_{n,k}(g_{q},\xi)_{u_m}}{S_{n,k}(g_{q},u_m)}$,
$m=n-k+1,...,n,$ which are linearly independent as functions of
$\xi$. Substituting these into (\ref{whit}), we obtain
\begin{equation}   \label{w1}
\sum_{i=1}^{n-k}\Big(\frac{S_{n,k}(g_{q},\xi)S_{n,k}(g_{r},u_i)-S_{n,k}(g_{r},\xi)S_{n,k}(g_{q},u_i)}
{(\xi-u_i)\cdot
u_i(u_i-1)(u_i-u_1)...\hat{i}...(u_i-u_{n-k})}u_{i,t_{s}}+$$
$$\frac{S_{n,k}(g_{r},\xi)S_{n,k}(g_{s},u_i)-S_{n,k}(g_{s},\xi)S_{n,k}(g_{r},u_i)}
{(\xi-u_i)\cdot
u_i(u_i-1)(u_i-u_1)...\hat{i}...(u_i-u_{n-k})}u_{i,t_{q}}+$$
$$\frac{S_{n,k}(g_{s},\xi)S_{n,k}(g_{q},u_i)-S_{n,k}(g_{q},\xi)S_{n,k}(g_{s},u_i)}
{(\xi-u_i)\cdot
u_i(u_i-1)(u_i-u_1)...\hat{i}...(u_i-u_{n-k})}u_{i,t_{r}}\Big)=0,
\end{equation}
\begin{equation}   \label{w2}
\sum_{i=1}^{n-k}\Big(\frac{S_{n,k}(g_{q},\xi)S_{n,k}(g_{r},u_i)-S_{n,k}(g_{r},\xi)S_{n,k}(g_{q},u_i)}
{(u_i-u_m)\cdot
u_i(u_i-1)(u_i-u_1)...\hat{i}...(u_i-u_{n-k})}u_{i,t_{s}}+$$
$$\frac{S_{n,k}(g_{r},\xi)S_{n,k}(g_{s},u_i)-S_{n,k}(g_{s},\xi)S_{n,k}(g_{r},u_i)}
{(u_i-u_m)\cdot
u_i(u_i-1)(u_i-u_1)...\hat{i}...(u_i-u_{n-k})}u_{i,t_{q}}+$$
$$\frac{S_{n,k}(g_{s},\xi)S_{n,k}(g_{q},u_i)-S_{n,k}(g_{q},\xi)S_{n,k}(g_{s},u_i)}
{(u_i-u_m)\cdot
u_i(u_i-1)(u_i-u_1)...\hat{i}...(u_i-u_{n-k})}u_{i,t_{r}}\Big)+$$
$$\frac{S_{n,k}(g_{q},\xi)S_{n,k}(g_{r},u_m)-S_{n,k}(g_{r},\xi)S_{n,k}(g_{q},u_m)}
{u_m(u_m-1)(u_m-u_1)...(u_m-u_{n-k})}u_{m,t_{s}}+$$
$$\frac{S_{n,k}(g_{r},\xi)S_{n,k}(g_{s},u_m)-S_{n,k}(g_{s},\xi)S_{n,k}(g_{r},u_m)}
{u_m(u_m-1)(u_m-u_1)...(u_m-u_{n-k})}u_{m,t_{q}}+$$
$$\frac{S_{n,k}(g_{s},\xi)S_{n,k}(g_{q},u_m)-S_{n,k}(g_{q},\xi)S_{n,k}(g_{s},u_m)}
{u_m(u_m-1)(u_m-u_1)...(u_m-u_{n-k})}u_{m,t_{r}}=0,
\end{equation}
where $m=n-k+1,...,n$. One can check straightforwardly that
(\ref{w2}) is equivalent to (\ref{syst1}) for $m=n-k+1$ and to
(\ref{syst2}) for $m=n-k+2,...,n$. Notice that the left hand side of
equation (\ref{w1}) is a polynomial in $\xi$ of degree $n-k-1$.
Evaluating this polynomial at $\xi=u_j,~j=1,...,n-k$ we obtain
(\ref{syst}). $\blacksquare$

{\bf Remark 3.} Given $t_{1},t_{2},t_{3},$ Theorem 2 yields a
3-dimensional system of the form (\ref{genern}) with $l=n+k$
equations possessing pseudopotential representation. Indeed, the
formulas (\ref{syst1}), (\ref{syst2}) give $3k$ linearly independent
equations if $q,r,s=1,2,3$. The formula (\ref{syst}) gives $n-k$
equations. On the other hand, one can construct exactly $k$ linear
combinations of equations (\ref{syst1}), (\ref{syst2}) with
$q,r,s=1,2,3$ such that derivatives of $u_i,~i=n-k+1,...,n$ cancel
out. Moreover, these linear combinations belong to the span of
equations (\ref{syst}). Therefore, there exist $(n-k)+3k-k=n+k$
linearly independent equations.

{\bf Remark 4.} In (\ref{syst}), (\ref{syst1}), (\ref{syst2}) we
have to assume $n\geq k+2$. Indeed, for $n=k+1$ we cannot construct
more then one pseudopotential and therefore there is no any system
of the form (\ref{genern}) associated with this case. However the
corresponding pseudopotential generates interesting integrable
(1+1)-dimensional systems of hydrodynamic type (see Section 5).
Probably these pseudopotentials for $k=0,1,...$ are also related to
some infinite integrable chains of the Benney type \cite{fer1,pav1}.

The system (\ref{syst})-(\ref{syst2}) possesses many conservation
laws of the hydrodynamic type. In particular, the following
statement can be verified by a straightforward calculation.

{\bf Proposition 9.} For any $r\ne s=0,1,...,n$ there exist $k$
conservation laws for the system (\ref{syst})-(\ref{syst2}) of the
form:
\begin{equation}   \label{conlaw}
\left(\frac{\Delta(g_r,h_1,...\hat{i}...h_k)}{\Delta(h_1,...,h_k)}\right)_{t_s}=
\left(\frac{\Delta(g_s,h_1,...\hat{i}...h_k)}{\Delta(h_1,...,h_k)}\right)_{t_r},
\end{equation}
where $i=1,...,k$. Here
$$\Delta(f_1,...,f_k)=\det\left(\begin{array}{ccc}f_1&...&f_k\\h_{1,u_{n-k+2}}&...&f_{k,u_{n-k+2}}
\\.........&...&.........\\ f_{1,u_n}&...&f_{k,u_n}
\end{array}\right).$$

Proposition 9 allows us to define functions $z_1,...,z_k$ such that
\begin{equation}   \label{z}
\frac{\Delta(g_r,h_1,...\hat{i}...h_k)}{\Delta(h_1,...,h_k)}=z_{i,t_r}
\end{equation}
for all $i=1,...,k$ and $r=0,1,...,n$.

Suppose $n\geq 3k$; then the system of the form (\ref{genern})
obtained from (\ref{syst}), (\ref{syst1}), (\ref{syst2}) with
$q,r,s=1,2,3$ consists of $3k$ equations (\ref{syst1}),
(\ref{syst2}) (they are equivalent to (\ref{conlaw})) and $n-2k$
equations of the form (\ref{syst}). Indeed, only $n-2k$ equations
(\ref{syst}) are linearly independent on (\ref{syst1}),
(\ref{syst2}). Expressing $u_1,...,u_{3k}$ in terms of
$z_{i,t_1},z_{i,t_2},z_{i,t_3},~i=1,...,k$ from (\ref{z}) and
substituting into $n-2k$ equations of the form (\ref{syst}), we
obtain a 3-dimensional system of $n-2k$ equations for $n-2k$
unknowns $z_1,...,z_k,u_{3k+1},...u_n$. This is a quasi-linear
system of the second order with respect to $z_i$ and of the first
order with respect to $u_j,$ whose coefficients depend on
$z_{i,t_1},z_{i,t_2},z_{i,t_3},~i=1,...,k$ and $u_{3k+1},...u_n$.
 It is clear that the general solution
of the system can be locally parameterized by $n-k$ functions in
two variables.

In the case $2k\leq n<3k$ the functions
$z_{i,t_1},z_{i,t_2},z_{i,t_3},~i=1,...,k$ are functionally
dependent. We have $3k-n$ equations of the form
$$R_i(z_{1,t_1},z_{1,t_2},z_{1,t_3},...,z_{k,t_1},z_{k,t_2},z_{k,t_3})=0,\qquad i=1,...,3k-n$$
and $n-2k$ second order quasi-linear equations. Totally we have
$(3k-n)+(n-2k)=k$ equations for $k$ unknowns $z_1,...,z_k$. It is
clear that the general solution of this system can be locally
parameterized by $n-k$ functions in two variables.

Suppose $n<2k$; then  we have  $n+k<3k,$ which means that $3k$ equations of
the form (\ref{syst1}), (\ref{syst2}) are linearly dependent. Probably in this case
 the general solution of the system can also be locally parameterized by
$n-k$ functions in two variables.

One of the most interesting cases is $n=3 k$, when we have a system of $k$ quasi-linear second order
equations for the functions  $z_1,...,z_k$. Consider the simplest case $k=1.$

{\bf Example 4.} In the case $n=3,\,k=1$ the formulas
(\ref{syst}), (\ref{syst1}) can be rewritten as follows. Let $h_{1}, g_{0},g_{1},g_{2}$
be linearly independent elements of ${\cal H}$. Denote by $B_{ij}$ the cofactors of the matrix
$$\left(\begin{array}{cccc}h_1&g_{0}&g_{1}&g_{2}\\ h_{1,u_{1}}&g_{0,u_{1}}&g_{1,u_{1}}&g_{1,u_{1}}
\\ h_{1,u_{2}}&g_{0,u_{2}}&g_{1,u_{2}}&g_{1,u_{1}}\\ h_{1,u_{3}}&g_{0,u_{3}}&g_{1,u_{3}}&g_{1,u_{3}}
\end{array}\right).
$$
Define vector fields $V_{i}$ by
$$
V_{1}=B_{22}\, \frac{\partial}{\partial  t_{0} }+B_{23}\, \frac{\partial}{\partial  t_{1} }+B_{24}\, \frac{\partial}{\partial  t_{2}
},
$$
$$
V_{2}=B_{32}\, \frac{\partial}{\partial  t_{0} }+B_{33}\, \frac{\partial}{\partial  t_{1} }+B_{34}\, \frac{\partial}{\partial  t_{2}
},
$$
$$
V_{3}=B_{42}\, \frac{\partial}{\partial  t_{0} }+B_{43}\, \frac{\partial}{\partial  t_{1} }+B_{44}\, \frac{\partial}{\partial  t_{2}
}.
$$
Then (\ref{syst1}) is equivalent to
\begin{equation}   \label{eq1}
V_{1}(u_{2})= V_{2}(u_{1}), \qquad V_{2}(u_{3})= V_{3}(u_{2}),  \qquad V_{3}(u_{1})=
V_{1}(u_{3}).
\end{equation}
Relation (\ref{syst}) leads to one more equation
\begin{equation}   \label{eq2}
u_{3} (u_{3}-1) (u_{1}-u_{2}) V_{1}(u_{2})+u_{1} (u_{1}-1) (u_{2}-u_{3})
V_{2}(u_{3})+u_{2} (u_{2}-1) (u_{3}-u_{1}) V_{3}(u_{1})=0.
\end{equation}
The conservation laws (\ref{conlaw}) have the form
$$
\left(\frac{ g_r }{h_1 }\right)_{t_s}=
\left(\frac{ g_s}{ h_1}\right)_{t_r}.
$$
Introducing $z$ such that $z_{t_{r}}=\frac{ g_r }{h_1 }$, we reduce (\ref{eq2}) to a quasi-linear equation of
the form \begin{equation}   \label{eqq}
\sum_{i,j} P_{i,j} (z_{t_0},z_{t_1},z_{t_2})\, z_{t_i,t_j}=0, \qquad
i,j=0,1,2.
\end{equation}
In the paper \cite{burferts} an inexplicit description of all
integrable equations (\ref{eqq}) was proposed. The equation
constructed above corresponds to the generic case in this
classification. Indeed, it depends on five essential parameters
$s_1,...,s_5$ which agrees with the results of \cite{burferts}.

For integer values of parameters $s_{i}$ equations (\ref{Darbu}),
(\ref{Hyper}) can be solved in elementary functions. This provides
simple examples of equations (\ref{eqq}) having pseudopotentials. In the most
degenerate case $s_{1}=\cdots=s_{5}=0$ one can choose $h_{1}=1,\,
g_{0}=u_{1},\, g_{1}=u_{2},\, g_{2}=u_{3}.$ The corresponding equation
(\ref{eqq}) is given by
$$
z_{t_2}(z_{t_2}-1)(z_{t_0}-z_{t_1}) z_{t_0t_1}+z_{t_0}(z_{t_0}-1)(z_{t_1}-z_{t_2}) z_{t_1t_2}+z_{t_1}(z_{t_1}-1)(z_{t_2}-z_{t_0})
z_{t_2t_0}=0.
$$
More general examples of equations
\begin{equation}   \label{bur}
P_{1}(z_{t_0},z_{t_1},z_{t_2})\, z_{t_0t_1}+P_{2}(z_{t_0},z_{t_1},z_{t_2})\,
z_{t_1t_2}+P_{3}(z_{t_0},z_{t_1},z_{t_2})\, z_{t_2t_0}=0
\end{equation}
correspond to
$s_{1}=s_{2}=s_{3}=0$. In this case one can choose $h=1$, $g_{0}=f(u_{1}),$
$g_{1}=f(u_{2}),$ $g_{2}=f(u_{3})$, where $f'(x)=x^{s_{4}} (x-1)^{s_{5}}$.
In the new variables $\bar u_{i}=f(u_{i})$ the
system (\ref{eq1}), (\ref{eq2}) is equivalent to a single equation of the
form
(\ref{bur}). One of the results of the paper \cite{burferts} is a
complete classification of equations (\ref{bur}) possessing a
pseudopotential representation. The above example seems to be the generic case in
this classification. $\blacksquare$

The system (\ref{syst})-(\ref{syst2}) has conservation laws different from (\ref{conlaw}).

{\bf Conjecture.} The system (\ref{syst}) -
(\ref{syst2}) possesses $n+1$  conservation laws of the
general form
$$A_{t_q}+B_{t_r}+C_{t_s}=0$$
additional to (\ref{conlaw}). This family of conservation
laws can be parameterized by elements from $\widehat{\cal H}=$ ${\cal
H}_{-2s_1,...,-2s_n,-2s_{n+1}-1,-2s_{n+2}-1}$ (cf. Proposition 7). This conjecture is
supported by some computer computations for small $n$ and $k$.

{\bf Remark 5.} Let us make in (\ref{pp2}) a change of variables  of
the form
\begin{equation}
\label{tr}\xi\to\frac{a\xi+b}{c\xi+d}, \qquad u_1\to\phi_1,...,u_n\to\phi_n,
\end{equation}
where $a,~b,~c,~d,~\phi_1,...,\phi_n$ are arbitrary functions in
$u_1,...,u_n$. After that we get  under the integral in (\ref{pp2})
an expression of the form
$$S(\xi)(\xi-\rho_1)^{-s_1-1}...(\xi-\rho_{n-k+1})^{-s_{n-k+1}-1}(\xi-\rho_{n-k+2})^{-s_{n-k+2}}...(\xi-\rho_n)^{-s_n}
(\xi-\rho_{n+1})^{-s_{n+1}-1}$$
$$
\times (\xi-\rho_{n+2})^{-s_{n+2}-1}(\xi-\rho_{n+3})^{s_1+...+s_{n+2}+1},
$$
where $S(\xi)$ is a polynomial in $\xi$ of degree $n-k$ and
$\rho_1,...,\rho_{n+3}$ are functions of $u_1,...,u_n$. Therefore
the numbers
\begin{equation}   \label{num}
\{-s_1-1,...,-s_{n-k+1}-1,-s_{n-k+2},...,-s_n,-s_{n+1}-1,-s_{n+2}-1,s_1+...+s_{n+2}+1\}
\end{equation}
play a symmetric role in the constructed pseudopotentials $A_{n,k}$.
Using transformations (\ref{tr}), one can choose any three of the
functions $\rho_1,...,\rho_{n+3}$ to be equal to $0,~1,~\infty$ and
the other $n$ functions to be equal to $u_1,...,u_n$ (cf.
\cite{odes}, Section 3). It would be interesting to study the
degenerate cases when some of the functions $\rho_i$ coincide (cf.
\cite{odpavsok}, Section 5).

The most symmetric case is given by
$$s_1=...=s_{n-k+1}=s_{n+1}=s_{n+2}=-\frac{k+1}{n+3},~~~s_{n-k+2}=...=s_n=\frac{n-k+2}{n+3}.$$
In this case all numbers (\ref{num}) are equal to
$-\frac{n-k+2}{n+3}$. Possibly for $n=3,~ k=1$ these values of
parameters correspond to pseudopotentials for integrable Lagrangians
of the form ${\cal L}(u_{x},u_{y},u_{z})$  \cite{ferhusts,odfer}
whereas for $n=5,~ k=3$ they are related to the integrable Hirota
type equations \cite{ferhadhus}.

{\bf Example 5.} Let $n=5$, $k=3$ and
$s_1=s_2=s_3=s_6=s_7=-\frac{1}{2},~s_4=s_5=\frac{1}{2}$. It turns out
that there exists a basis $g_1,~g_2,~g_3,~h_1,~h_2,~h_3$ in
$\cal H$ such that
$$\Delta(g_1,h_1,h_2)=\Delta(g_3,h_2,h_3),$$
\begin{equation}
\label{dep} \Delta(g_2,h_2,h_3)=\Delta(g_1,h_3,h_1),
\end{equation}
$$\Delta(g_3,h_3,h_1)=\Delta(g_2,h_1,h_2).$$
Indeed, the system (\ref{dep}) is a consequence of equations
$$g_1h_{1,u_4}-h_1g_{1,u_4}+g_2h_{2,u_4}-h_2g_{2,u_4}+g_3h_{3,u_4}-h_3g_{3,u_4}=0,$$
\begin{equation}
\label{dep1}
g_1h_{1,u_5}-h_1g_{1,u_5}+g_2h_{2,u_5}-h_2g_{2,u_5}+g_3h_{3,u_5}-h_3g_{3,u_5}=0,
\end{equation}
$$g_{1,u_4}h_{1,u_5}-h_{1,u_4}g_{1,u_5}+g_{2,u_4}h_{2,u_5}-h_{2,u_4}g_{2,u_5}+g_{3,u_4}h_{3,u_5}-h_{3,u_4}g_{3,u_5}=0.$$
Consider the system consisting of equations (\ref{dep1}) and all its
first and second derivatives with respect to $u_1,...,u_5$. Note
that differentiating (\ref{dep1}), we eliminate second derivatives
of $h_i$ and $g_i$ by (\ref{Darbu}), (\ref{Hyper}). One can check
that this system is invariant with respect to the derivations by
$u_1,...,u_5$. At a fixed generic point $u_1^0,...,u_5^0$ the system
can be regaded as an algebraic variety for the values of $g_i,~h_i$
and their first derivatives. It can be checked that this variety
consists of several components and the maximal dimension of the
component equals 24. Since the vector fields
$\frac{\partial}{\partial u_i}$ are tangent to this variety, any its
point  considered as the initial data defines the solutions
$g_i,~h_i$ of (\ref{Darbu}), (\ref{Hyper}) such that the
corresponding point belongs to the variety for any values of
$u_1,...,u_5.$ It is possible to check that there exists an
algebraic component of dimension 21 of the variety such that the
Wronskian of $g_i,~h_i$ at $u_1^0,...,u_5^0$ is non-zero.

Proposition 9 and equations (\ref{dep}) allow us to define a
function $z$ such that
$$z_{t_1,t_1}=\frac{\Delta(g_1,h_2,h_3)}{\Delta(h_1,h_2,h_3)},~z_{t_2,t_2}=\frac{\Delta(g_2,h_3,h_1)}{\Delta(h_1,h_2,h_3)},~
z_{t_3,t_3}=\frac{\Delta(g_3,h_1,h_2)}{\Delta(h_1,h_2,h_3)},$$
$$z_{t_1,t_2}=\frac{\Delta(g_2,h_2,h_3)}{\Delta(h_1,h_2,h_3)}=\frac{\Delta(g_1,h_3,h_1)}{\Delta(h_1,h_2,h_3)},~
z_{t_2,t_3}=\frac{\Delta(g_3,h_3,h_1)}{\Delta(h_1,h_2,h_3)}=\frac{\Delta(g_2,h_1,h_2)}{\Delta(h_1,h_2,h_3)},$$
$$z_{t_3,t_1}=\frac{\Delta(g_1,h_1,h_2)}{\Delta(h_1,h_2,h_3)}=
\frac{\Delta(g_2,h_1,h_2)}{\Delta(h_1,h_2,h_3)}.$$ It terms of this
function we can rewrite the system (\ref{syst}), (\ref{syst1}),
(\ref{syst2}) as a single equation of the form (\ref{hir}).
Integrable systems of this form were studied in \cite{ferhadhus}.
The pseudopotentials considered above correspond to the generic
integrable system of this form.

{\bf Remark 6.} It is easy to see that the group $SP_6$ acts on the
set of bases in $\cal H$ satisfying (\ref{dep1}). This agrees with
the result of \cite{ferhadhus} that this group acts on the set of
integrable equations of the form (\ref{hir}).$\blacksquare$

\section{Integrable (1+1)-dimensional systems of hydrodynamic type}

In this section we consider integrable  (1+1)-dimensional
hydrodynamic type systems (\ref{gidra}) constructed in terms of
generalized hypergeometric functions. These systems appear as the
so-called hydrodynamic reductions of pseudopotentials $A_{n,k}$ (see
the next Section). By integrability we mean the existence of
infinite number of hydrodynamic commuting flows and conservation
laws. It is known \cite{tsar} that this is equivalent to the
following relations for the velocities $v^{i}(r^1,...,r^N)$:
\begin{equation}   \label{semiham}
\partial_{j}\frac{\partial_{i} v^{k}}{v^{i}-v^{k}}=\partial_{i}\frac{\partial_{j}
v^{k}}{v^{j}-v^{k}}, \qquad  i\ne j\ne k, \qquad
\end{equation}
Here $\partial_{\alpha}=\frac{\partial}{\partial r^{i}}, \,
\alpha=1,\dots,N$. The system (\ref{gidra}) is called {\it semi-Hamiltonian}
if conditions (\ref{semiham}) hold.

The main geometrical object related to a semi-Hamiltonian system
(\ref{gidra}) is a diagonal metric $g_{kk},\,k=1,\dots,N$, where
\begin{equation}   \label{metrik}
\frac{1}{2}\partial_{i} \log{g_{kk}} = \frac{\partial_{i}
v^{k}}{v^{i}-v^{k}}, \qquad i\ne k.
\end{equation}
In view of (\ref{semiham}), the overdetermined system (\ref{metrik}) is
compatible and the function $g_{kk}$ is defined up to arbitrary factor
$\eta_{k}(r^{k})$. The metric $g_{kk}$ is called the {\it metric associated to (\ref{gidra}).}
It is known that two hydrodynamic type systems are compatible
iff they possess a common associated metric \cite{tsar}.

A diagonal metric $g_{kk}$
is called a {\it metric of Egorov type} if for any $i,j$
\begin{equation}   \label{egor}
\partial_i g_{jj}=\partial_j g_{ii}.
\end{equation}
Note that if a Egorov-type metric associated with a
hydrodynamic-type system of the form (\ref{gidra}) exists, then it
is unique. For any Egorov's metric there exists a potential $G$ such
that $g_{ii}=\partial_i G$. Semi-Hamiltonian systems possessing
associated metrics of Egorov type play important role in the theory
of WDVV associativity equations and in the theory of Frobenious
manifolds \cite{dub,kr,pavts}.

 Let
$w(r^1,...,r^N),~\xi_1(r^1,...,r^N),...,\xi_N(r^1,...,r^N)$ be a
solution of (\ref{gibtsar1}). It can be easily verified that this
system is in involution and therefore its solution admits a local
parameterization by $2N$ functions of one variable.  Let
$u_1(r^1,...,r^N),...,u_n(r^1,...,r^N)$ be a set of solutions of the
system (\ref{u1}). It is easy to verify that this system is in
involution and therefore has an one-parameter family of solutions
for fixed $\xi_{i},w$.

Consider the following system
\begin{equation}\label{gidragen}
r^i_t=\frac{S_{n,k}(g_1,\xi_i)}{S_{n,k}(g_2,\xi_i)}r^i_x,
\end{equation}
where $g_{1},g_{2}$ are linearly independent solutions of
(\ref{Darbu}), (\ref{Hyper}), the polynomials $S_{n,k}, \,k>0$ are
defined by (\ref{polgen}), and $S_{n,0}=S_{n}$ (see formula
(\ref{pol0})).

{\bf Theorem 3.} The system (\ref{gidragen}) is semi-Hamiltonian. The
associated metric is given by
$$g_{ii}=S_n(g_2,\xi_i)^2(\xi_i-u_1)^{-2s_1-2}\cdots (\xi_i-u_n)^{-2s_n-2}
\xi_i^{-2s_{n+1}-1}(\xi_i-1)^{-2s_{n+2}-1}\partial_iw$$ for $k=0,$
and by
$$g_{ii}=S_{n,k}(g_2,\xi_i)^2(\xi_i-u_1)^{-2s_1-2}\cdots (\xi_i-u_{n-k+1})^{-2s_{n-k+1}-2}\times
$$$$(\xi_i-u_{n-k+2})^{-2s_{n-k+2}}\cdots
(\xi_i-u_n)^{-2s_n}\xi_i^{-2s_{n+1}-1}(\xi_i-1)^{-2s_{n+2}-1}\partial_iw$$
for $k>0$.

{\bf Proof.} Substituting the expression for the metric into
(\ref{metrik}), where $v^i$ are specified by (\ref{gidragen}), one
obtains the identity by virtue of (\ref{gibtsar1}), (\ref{u1}).
$\blacksquare$

{\bf Remark 7.} The system (\ref{gidragen}) does not possess the
associated metric of the Egorov type in general. However, for very
special values of the parameters $s_{i}$ in (\ref{Darbu}),
(\ref{Hyper}) there exists $g_2\in\cal H$ such that the metric is of
the Egorov type for all solutions of the system (\ref{gibtsar1}),
(\ref{u1}). For instance, if the defect $k$ equals zero, then this
happens exactly in the following cases:

$s_{i}=0$ for all $i$;

$s_{l}=-1$ for some $l$ and $s_{i}=0$ for $i\ne l$;

$s_{l}=-\frac{1}{2}$ for some $l$ and $s_{i}=0$ for $i\ne l$;

$s_{j}=s_{l}=-\frac{1}{2}$ for some $j\ne l$ and $s_{i}=0$ for
$i\ne j, i\ne l$.

{\bf Proposition 10.} Suppose that a solution $\xi_{i},w$ of
(\ref{gibtsar1}) and solutions $u_1,...,u_n$ of  (\ref{u1}) are
fixed. Then the hydrodynamic type systems
\begin{equation}\label{gidragencom}
r^i_{t_1}=\frac{S_{n,k}(g_1,\xi_i)}{S_{n,k}(g_3,\xi_i)}r^i_x,~~~~~~~r^i_{t_2}=\frac{S_{n,k}(g_2,\xi_i)}{S_{n,k}(g_3,\xi_i)}r^i_x
\end{equation} are compatible for all $g_1,~g_2$.

{\bf Proof.} Indeed, the metric associated with (\ref{gidragen})
does not depend on $g_2$. Therefore the systems (\ref{gidragencom})
has a common metric depending on $g_3$ and on solutions of
(\ref{gibtsar1}), (\ref{u1}). $\blacksquare$

{\bf Remark 8.} One can also construct some compatible systems of
the form (\ref{gidragencom}) using Proposition 5. Set
$g_2=Z(u_1,...,u_n,u_{n+1})$ in (\ref{gidragencom}). Here $u_{n+1}$
is an arbitrary solution of (\ref{u1}) distinct from $u_1,...,u_n$.
It is clear that the flows (\ref{gidragencom}) are compatible for
such $g_2$ and any $g_1\in \cal H$. Moreover, Proposition 5 implies
that the flows (\ref{gidragencom}) are compatible if we set
$g_1=Z(u_1,...,u_n,u_{n+1})$, $g_2=Z(u_1,...,u_n,u_{n+2})$ for two
arbitrary solutions $u_{n+1},~u_{n+2}$ of (\ref{u1}).

All members of the hierarchy constructed in Proposition 10 possess a
dispersionless Lax representation (\ref{maineq}) with common
$L(p,r^1,...,r^N)$.  Define a function $L(\xi,r^1,...,r^N)$ by the
following system
\begin{equation}\label{L1}
\partial_iL=\frac{\xi(\xi-1)\partial_iw\,L_{\xi}}{\xi-\xi_i},\qquad
i=1,...,N.
\end{equation}
Note that the system (\ref{L1}) is in involution and therefore the
function $L$ is defined uniquely up to inessential transformations
$L\to\lambda(L)$. To find the function $L(p,r^1,...,r^N)$ one has to
express $\xi$ in terms of $p$  by (\ref{psdef0}) for $k=0$ or by
(\ref{psdef}) for $k>0$.

{\bf Proposition 11.} Let  $u_{1}, \dots, u_{n}$ be arbitrary
solution of (\ref{u1}). Then system (\ref{gidragen}) admits the
dispersionless Lax representation (\ref{maineq}),
 where $A=A_{n,k}$ is defined by (\ref{psdef0}) for $k=0$ and by
 (\ref{psdef}) for $k>0$.

{\bf Proof.} Substituting $A=A_{n,k}$ defined by (\ref{psdef0}) for
$k=0$ and by
 (\ref{psdef}) for $k>0$ into
(\ref{maineq}) and calculating $L_t$ by virtue of (\ref{gidragen})
we arrive to the expression
$$\partial_iL=\frac{\partial_iP_{n,k}(g_2,\xi)\cdot S_{n,k}(g_1,\xi_i)-\partial_iP_{n,k}(g_1,\xi)\cdot S_{n,k}(g_2,\xi_i)}
{P_{n,k}(g_2,\xi)_{\xi}\cdot
S_{n,k}(g_1,\xi_i)-P_{n,k}(g_1,\xi)_{\xi}\cdot
S_{n,k}(g_2,\xi_i)}L_{\xi}.$$ Taking into account the equation
$P_{n,k}(g_i,\xi)_{\xi}=S_{n,k}(g_i,\xi)(\xi-u_1)^{-s_1-1}...(\xi-u_{n-k+1})^{-s_{n-k+1}-1}
(\xi-u_{n-k+2})^{-s_{n-k+2}}...(\xi-u_n)^{-s_n}\xi^{-s_{n+1}-1}(\xi-1)^{-s_{n+2}-1}$
and writing down $P_{n,k}(g_i,\xi)_{u_m}$ in terms of
$P_{n,k}(g_1,\xi)_{u_{n-k+1}},...,P_{n,k}(g_1,\xi)_{u_n}$ by
(\ref{pr4}), (\ref{pr3}), one can readily check this equation. $\blacksquare$

Let the function
$\xi(L,r^1,...,r^N)$ be inverse to $L(\xi,r^1,...,r^N).$
It is easy to check that
$u=\xi(L,r^1,...,r^N),$ where $L$ plays a role of arbitrary parameter, satisfies
(\ref{u1}).

As usual, the Lax representation defines conserved  densities, common
for the whole hierarchy, by formula (\ref{con}). Since our
definition of $A_{n,k}$ is parametric, we can reformulate this fact
as

{\bf Proposition 12.} Suppose (\ref{gidragen}) is defined by
solutions $u_{1},\dots, u_{n}$ of the system (\ref{u1}). Let $U$ be
any solution of (\ref{u1}). Then
$$
\frac{\partial}{\partial t}P_{n,k}\Big(g_2(u_{1},u_{1},\dots,
u_{n}),U\Big) =\frac{\partial}{\partial x}
P_{n,k}\Big(g_1(u_{1},u_{1},\dots, u_{n}),U\Big)
$$
is a conservation law for (\ref{gidragen}).

Since the generic solution $U$ depends on a parameter, we have
constructed an one-parametric family of common conservation laws for
our hierarchy (\ref{gidragen}) of hydrodynamic type systems.

\section{Hydrodynamic reductions and integrability}

In this section we show that integrable (1+1)-dimensional systems
constructed in Section 5 define hydrodynamic reductions for
pseudopotentials and 3-dimensional systems from Sections 3 and 4.

Following \cite{ferhus1, pav2, odpavsok}, we give a definition of
integrability for equations (\ref{maineq}), (\ref{psi}) and
(\ref{genern}) in terms of hydrodynamic reductions.

Suppose there exists a pair of compatible semi-Hamiltonian
hydrodynamic-type systems of the form
\begin{equation}\label{dred}
r_{t_1}^{i}=v_1^{i}(r^1,...,r^N)r_{x}^{i}, \qquad r_{t_2}^{i}=v_2^{i}(r^1,...,r^N)r_{x}^{i}
\end{equation}
and functions
$u_i=u_i(r^1,...,r^N)$ such that these functions satisfy (\ref{genern}) for any
solution of (\ref{dred}).  Then
(\ref{dred}) is called {\it a hydrodynamic reduction} for
(\ref{genern}).

{\bf Definition 1} \cite{ferhus1}. A system of the form (\ref{genern}) is called
{\it integrable} if equation (\ref{maineq}) possesses
sufficiently many hydrodynamic reductions for each $N\in\N$.
"Sufficiently many" means that the set of hydrodynamic reductions
can be locally parameterized by $2N$ functions of one variable. Note
that due to gauge transformations $r^i\to\lambda_i(r^i)$ we have $N$
essential functional parameters for hydrodynamic reductions.

Suppose there exists a semi-Hamiltonian
hydrodynamic-type system (\ref{gidra}) and functions
$u_i=u_i(r^1,...,r^N)$, $L=L(p,r^1,...,r^N)$ such that these
functions satisfy dispersionless Lax equation (\ref{maineq}) for any solution $
r^1(x,t),...,r^N(x,t)$ of the system (\ref{gidra}). Then
(\ref{gidra}) is called {\it a hydrodynamic reduction} for
  (\ref{maineq}).

{\bf Definition 2} \cite{odpavsok}. A dispersionless Lax equation (\ref{maineq}) is called
{\it integrable} if equation (\ref{maineq}) possesses
sufficiently many hydrodynamic reductions for each $N\in\N$.

We also call the corresponding pseudopotential $A(p,u_1,...,u_n)$ integrable.

{\bf Example 6.} Let us show that $A=\ln(p-u)$ is integrable. Let
$w(r^1,...,r^N)$, $p_i(r^1,...,r^N)$, $i=1,...,N$ be an arbitrary
solution of the following system (the so-called Gibbons-Tsarev
system \cite{Gibt})
\begin{equation}\label{gibtsar}
\partial_j\xi_i=\frac{\partial_jw}{\xi_j-\xi_i},\qquad \partial_{ij}w=\frac{2\partial_iw\partial_jw}{(\xi_i-\xi_j)^2}, \qquad i,
j=1,...,N,~~~i\ne j.
\end{equation}
It is easy to verify that this system  is in involution and
therefore its general solution admits a local parameterizations by
$2N$ functions of one variable. Define a function $L(p,r^1,...,r^N)$
by the following system
\begin{equation}\label{L}
\partial_iL=\frac{\partial_iw\,L_p}{p-\xi_i}, \qquad i=1,...,N.
\end{equation}
This system  is in involution and therefore
defines the function $L$ uniquely up to inessential transformations
$L\to\lambda(L)$.
 Finally, let $u(r^1,...,r^N)$ be a solution of the
system
\begin{equation}\label{u}
\partial_iu=\frac{\partial_iw}{\xi_i-u},~~~i=1,...,N.
\end{equation}
It is easy to check that the system (\ref{u}) is in involution.  It follows from
 (\ref{gibtsar}), (\ref{L}), (\ref{u}) that the system
\begin{equation}\label{gidraln}
r^i_t=\frac{1}{\xi_i-u}r^i_x
\end{equation}
is a hydrodynamic reduction of equation (\ref{maineq}) with $A=\ln(p-u)$.

{\bf Remark 9.} The standard  form for the Gibbons-Tsarev system
\cite{pav2} related to hydrodynamic reductions is given by
$$
\partial_i \xi_{j}=F(\xi_{i},\xi_{j}, u_{1},\dots,u_{n})\, \partial_i u_{n},
\qquad
\partial_i \partial_{j} u_{n}= H(\xi_{i},\xi_{j}, u_{1},\dots,u_{n}) \,  \partial_i u_{n}\partial_j
u_{n},\qquad i\ne j
$$
$$
\partial_i u_{l}=G_l(\xi_{i}, u_{1},\dots,u_{n})\, \partial_i
u_{n},\qquad l<n.
$$
Here  $i,j=1,...,N, \quad$ $u_{l}(r^{1},\dots,r^{N})$ are the
functions, which define the reduction, and \linebreak
$\xi_{i}(r^{1},\dots,r^{N})$ are some auxiliary functions. To bring
(\ref{gibtsar}), (\ref{u}) to this form, one has to eliminate the
additional unknown $w$. The result is given by
\begin{equation}\label{gibtsar2}
\partial_j\xi_i=\frac{\xi_{i}-u}{\xi_j-\xi_i} \partial_j u,\qquad
\partial_i \partial_{j} u=\frac{\xi_i+\xi_j-2 u}{(\xi_i-\xi_j)^2} \, \partial_i u\partial_j
 u.
\end{equation}
In this case $n=1, u_{1}=u.$
There is the following generalization of (\ref{gibtsar2})
to the case of arbitrary polynomial $P(x)=a_3 x^3+a_2
x^2+a_1 x+ a_0$ and arbitrary $n$:

\begin{equation}\label{gibtsar3}\begin{array}{c}
\displaystyle \frac{u_1-\xi_i}{P(u_1)}\partial_iu_1=...=\frac{u_n-\xi_i}{P(u_n)}\partial_iu_n,~~~i=1,...,N,
\\[6mm]
\displaystyle \partial_{ij}u_n=\frac{K_2(\xi_i,\xi_j) u_n^2+K_1(\xi_i,\xi_j) u_n+
K_0(\xi_i,\xi_j) } {P(u_n)(\xi_i-\xi_j)^2}\,
\partial_i u_n\partial_j u_n,\\[6mm]
\displaystyle \partial_i\xi_j=\frac{P(\xi_j)(u_n-\xi_i)}{P(u_n)(\xi_i-\xi_j)}\partial_i u_n,~~~i,j=1,...,N,~~~i\ne
j,\end{array}
\end{equation}
 where
$$\begin{array}{c}
K_2(\xi_i,\xi_j)=2 a_3 (\xi_i-\xi_j)^2,\\[4mm]
K_1(\xi_i,\xi_j)=-a_3 (\xi_i^2\xi_j+\xi_i\xi_j^2)+a_2
(\xi_i^2+\xi_j^2-4\xi_i\xi_j) - a_1 (\xi_i+\xi_j)-2
a_0,\\[4mm]
K_0(\xi_i,\xi_j)= 2 a_3 \xi_i^2 \xi_j^2+ a_2
(\xi_i^2\xi_j+\xi_i\xi_j^2)+ a_1 (\xi_i^2+\xi_j^2)+a_0
(\xi_i+\xi_j).
\end{array}
$$
Using transformations of the form $\displaystyle
u_i\to\frac{au_i+b}{cu_i+d}$, $\displaystyle
\xi_i\to\frac{a\xi_i+b}{c\xi_i+d},$ one can put the
polynomial $P$ to one of the canonical forms: $P(x)=x(x-1)$, $P(x)=x$, or
$P(x)=1$. If $P(x)=1,$ then (\ref{gibtsar3}) with $n=1$ coincides with (\ref{gibtsar2}).
Formulas (\ref{gibtsar1}), (\ref{u1}) are equivalent to (\ref{gibtsar3}), where $P(x)=x(x-1)$.

{\bf Definition 3.} Two integrable pseudopotentials $A_1,~A_2$ are
called {\it compatible} if the system
$$L _{t_1}=\{L ,A_1 \}, \qquad L _{t_2}=\{L ,A_2\}$$
possesses sufficiently many  hydrodynamic reductions (\ref{dred})
 for each
$N\in\N$.

If $A_1,~A_2$ are compatible, then $A=c_1A_1+c_2A_2$ is
integrable for any constants $c_1$, $c_2$. Indeed, the system
$$r_{t}^{i}=(c_1 v_1^{i}(\mathbf{r})+c_2 v_2^i(\mathbf{r}))\,r_{x}^{i}$$ is a
hydrodynamic reduction of (\ref{maineq}).

{\bf Example 7.} The functions $A_1=\ln(p-u_1)$ and
$A_2=\ln(p-u_2)$ are compatible. Moreover,
$A=c_1\ln(p-u_1)+...+c_n\ln(p-u_n)$ is integrable for any constants
$c_1,...,c_n$.
Indeed, let $w$, $p_i$ satisfy (\ref{gibtsar}) and $u_1$, $u_2$ be
two different solutions of (\ref{u}). It is easy to check that the
corresponding flows are compatible by virtue of (\ref{gibtsar}),
(\ref{L}), (\ref{u}).

{\bf Definition 4.} By 3-dimensional system associated with compatible
functions $A_1,~A_2$ we mean the system of the
form (\ref{genern}) equivalent to compatibility conditions for the
system
\begin{equation}\label{sys}
\psi_{t_{2}}=A_{1}(\psi_{t_{1}},u_1,...,u_n), \qquad
\psi_{t_{3}}=A_{2}(\psi_{t_{1}},u_1,...,u_n). \end{equation}

It is clear that any system associated with a pair of compatible
functions possesses sufficiently many hydrodynamic
reductions and therefore it is integrable in the sense of Definition 1.

{\bf Example 8.} Let $A_1=\ln(p-u)$ and $A_2=\ln(p-v)$. The associated
3-dimensional system has the form
$$u_{t_3}=v_{t_2},~~~v_{t_1}-vu_{t_3}=u_{t_1}-uv_{t_2}.
$$

The following statement is a reformulation of Proposition 11.

{\bf Theorem 4.} The system (\ref{gidragen}) is a hydrodynamic
reduction of the pseudopotential $A_{n,k}$ defined by (\ref{psdef0})
if $k=0$ and by (\ref{psdef}) if $k>0$. Recall that we use the
notation $S_n\equiv S_{n,0},~ A_n\equiv A_{n,0},~ P_n\equiv
P_{n,0}$.

{\bf Proposition 13.} Suppose $g_1,g_2,g_3,h_1,...,h_k\in {\cal H}$
are linearly independent. Define pseudopotentials $A_1$, $A_2$ by
$$A_1=P_{n,k}(g_1,\xi),\qquad A_2=P_{n,k}(g_2,\xi), \qquad p=P_{n,k}(g_3,\xi).$$
Then $A_1$ and $A_2$ are compatible.

{\bf Proof.} Note that the system (\ref{gibtsar1}), (\ref{L1}) does
not depend on $g_1,g_2,g_3$ and therefore we have a family of
functions $L, \xi_i, u_i$ which give hydrodynamic reduction of the
form (\ref{gidragen}) for both $A_1$ and $A_2$. Moreover, according
to Proposition 10 the flows
$$r^i_{t_1}=\frac{S_{n,k}(g_1,\xi_i)}{S_{n,k}(g_3,\xi_i)}r^i_x, \qquad
r^i_{t_2}=\frac{S_{n,k}(g_2,\xi_i)}{S_{n,k}(g_3,\xi_i)}r^i_x$$ are
compatible. $\blacksquare$

{\bf Remark 10.} This result implies that 3-dimensional hydrodynamic
type systems constructed in Sections 4, 5 possess sufficiently many
hydrodynamic reductions.

{\bf Remark 11.} Using proposition 5, one can construct compatible
pseudopotentials depending on different number of $u_i$. Indeed, let
$g_1,g_3,h_1,...,h_k\in {\cal H}$ and $g_2=Z(u_1,...,u_n,u_{n+1})$.
Then $A_2$ depends on $u_1,...,u_n,u_{n+1}$ and $A_1$ depends on
$u_1,...,u_n$ only.

\section{Conclusion}

All known integrable pseudopotentials $A(p,u_1,...,u_n)$ satisfy the
property
$$P\left(\frac{A_{ppp}}{A_{pp}^2},A_p\right)=0,$$
where $P(x,y)$ is a polynomial in $x,~y$ with coefficients depending
on $u_1,...,u_n$. In this sense any pseudopotential $A$ is
associated with the algebraic curve ${\cal E}=\{(x,y)\in\C^2;~
P(x,y)=0\}$. Moreover, compatible pseudopotentials are associated to
isomorphic curves. If a 3-dimensional dispersionless system is
constructed by two compatible pseudopotentials, then this curve is
isomorphic to the so-called spectral curve (see \cite{ferhus1}) of
the system. In this paper we have constructed a wide class of
integrable pseudopotentials associated with rational curves. We
believe that all pseudopotentials associated with rational curves
can be obtained as a limit from our pseudopotentials. We are going
to describe all such limits in a separate paper.

It is known \cite{kr4} that pseudopotentials associated with curves
of higher genus also exist. It is likely that one can describe all
pseudopotentials associated with the elliptic curve in a similar
manner to the way we have done the rational case in this paper. We
are going to consider this problem in the next paper.

\vskip.3cm \noindent {\bf Acknowledgments.} Authors thank B.A.
Dubrovin, E.V. Ferapontov, I.M. Krichever, O.I. Mokhov, and M.V. Pavlov for fruitful
discussions.  V.S. thanks IHES and A.O. thanks MPIM and IHES for
hospitality and financial support. V.S. was partially supported by
the RFBR grants 08-01-461 and NS 3472.2008.2.


\begin{thebibliography}{10}

\bibitem{zahshab} \emph{V.E. Zakharov, A.B. Shabat}, \newblock{Integration of non-linear equations of mathematical physics by
the inverse scattering method},  Func. Anal. and Appl.
\textbf{13}(3) (1979) 13--22.

\bibitem{kr4}  \emph{I.M. Krichever}, \newblock{The $\tau$-function of the universal
Whitham hierarchy, matrix models and topological field theories},
\newblock{Comm. Pure Appl. Math.}, {\bf 47} (1994), no. 4, 437--475.

\bibitem{dub} \emph{B.A. Dubrovin}, \newblock  Geometry of 2D
topological field theories. \newblock{\em In Integrable Systems and
Quantum Groups},
\newblock Lecture Notes in Math.
{\bf 1620} (1996), 120--348.

\bibitem{odsok} \emph{A. Odesskii, V. Sokolov}, \newblock{On (2+1)-dimensional
hydrodynamic-type systems possessing pseudopotential with movable
singularities}, \newblock{\em  to appear in Func. Anal. Appl.}

\bibitem{odes} \emph{A.V. Odesskii},
\newblock{A family of (2+1)-dimensional hydrodynamic-type systems possessing pseudopotential},
arXiv:0704.3577v3 [math. AP], to appear in Selecta Mathematica.


\bibitem{pav4} \emph{M.V. Pavlov}, \newblock Classification of the Egorov hydrodynamic
chains. Theor. Math. Phys. \textbf{138} No. 1 (2004) 55-71.


\bibitem{odpavsok} \emph{A. Odesskii, M.V. Pavlov and V.V. Sokolov}, \newblock Classification of
integrable Vlasov-type equations,  arXiv:0710.5655, Theor. Math.
Phys. \textbf{154}(2)(2008) 209-219.

\bibitem{trfun} \emph{H. Bateman, A. Erdelyi}, \newblock{Higher
transcendental functions}, 1953.

\bibitem{gel} \emph{I.M. Gelfand, M.I. Graev, V.S. Retakh},  \newblock{General hypergeometric
systems of equations and series of hypergeometric type}, Russian
Math. Surveys 47 (1992), no. 4, 1--88

\bibitem{odfer} \emph{ E.V. Ferapontov, A.V. Odesskii}, Integrable Lagrangians and modular
forms, arXiv:0707.3433.

\bibitem{burferts} \emph{P.A. Burovskiy, E.V. Ferapontov and S.P. Tsarev},
\newblock{Second order quasilinear PDEs and conformal structures in projective space},
arXiv:0802.2626 [nlin. SI].

\bibitem{dorfer} \emph{W. F. Ames, R. L. Anderson, V. A. Dorodnitsyn, E. V. Ferapontov,
R. K. Gazizov, N. H. Ibragimov, S. R. Svirshchevskii}, CRC handbook
of Lie group analysis of differential equations. Vol. 1. Symmetries,
exact solutions and conservation laws. CRC Press, Boca Raton, FL,
1994.

\bibitem{ferhadhus} \emph{E.V. Ferapontov, L. Hadjikos, K.R. Khusnutdinova}, \newblock Integrable
equations of the dispersionless Hirota type and hypersurfaces in the
Lagrangian Grassmannian, arXiv:0705.1774.


\bibitem{Gibt} \emph{J. Gibbons, S.P. Tsarev}, \newblock Reductions of
Benney's equations, Phys. Lett. A, \textbf{211 }(1996) 19-24.
\emph{J. Gibbons, S.P. Tsarev}, \newblock Conformal maps and
reductions of the Benney equations, Phys. Lett. A, \textbf{258}
(1999) 263-270.

\bibitem{pav2} \emph{M.V. Pavlov}, \newblock Algebro-geometric
approach in the theory of integrable hydrodynamic-type systems.
Comm. Math. Phys., \textbf{272}(2) (2007) 469-505.

\bibitem{kr}  \emph{A. A. Akhmetshin, I. M. Krichever, Y. S. Volvovski},
 A generating formula for solutions of associativity equations. Russian Math. Surveys {\bf 54} (1999), no. 2,
 427--429.

\bibitem{ferhus1} \emph{E.V. Ferapontov, K.R. Khusnutdinova}, \newblock On
integrability of (2+1)-dimensional quasilinear systems, Comm. Math.
Phys.
\textbf{248} (2004) 187-206, \emph{E.V. Ferapontov, K.R. Khusnutdinova}, %
\newblock The characterization of 2-component (2+1)-dimensional integrable
systems of hydrodynamic type, J. Phys. A: Math. Gen. \textbf{37}(8)
(2004) 2949--2963.

\bibitem{fer1} \emph{E.V. Ferapontov, D.G. Marshal},
\newblock{Differential-geometric approach to the integrability of hydrodynamic chains: the Haanties
tensor},
\newblock{Math. Ann. \textbf{339}(1), (2007) 61--99}.

\bibitem{pav1} \emph{M.V. Pavlov}, \newblock{Classification of integrable hydrodynamic
chains and generating functions of conservation laws}, J. Phys. A:
Math. Gen. \textbf{39}(34) (2006) 10803--10819.


\bibitem{ferhusts}  \emph{ E.V. Ferapontov,  K.R. Khusnutdinova and S.P. Tsarev}, \newblock
On a class of three-dimensional integrable
Lagrangians, Comm. Math. Phys. {\bf 261}, no. 1 (2006)  225--243.


\bibitem{tsar} \emph{S.P. Tsarev}, \newblock On Poisson brackets and
one-dimensional Hamiltonian systems of hydrodynamic type, Soviet
Math. Dokl., \textbf{31} (1985) 488--491. \emph{S.P. Tsarev},
\newblock The geometry of Hamiltonian systems of hydrodynamic type.
The generalized hodograph method, Math. USSR Izvestiya, \textbf{37}
No. 2 (1991) 397--419. 1048--1068.

\bibitem{pavts} \emph{M.V. Pavlov, S.P. Tsarev}, \newblock Tri-Hamiltonian structures of the Egorov
systems of hydrodynamic type. Func. Anal. and Appl. , \textbf{37}(1)
(2003) 32-45.

\end{thebibliography}
\end{document}